\documentclass[aps,prl,twocolumn,superscriptaddress]{revtex4-2}

\usepackage{graphicx}
\usepackage{dcolumn}
\usepackage{bm}
\usepackage{subfigure}
\usepackage{mathrsfs}
\usepackage[T1]{fontenc}
\usepackage{multirow}
\usepackage{amsmath}

\usepackage[colorlinks,
linkcolor={red!60!black!},
anchorcolor={green!80!black!},
urlcolor={blue!80!black!},
citecolor={blue!50!black!}]{hyperref}
\usepackage[table]{xcolor}

\gdef\N3LOlnl{N$^3$LO+3N$_{\rm lnl}$}

\gdef\BE2{$B({\rm E}2)$}

\newcommand{\A}[2]{$^{#1}$#2}

\begin{document}

\title{Ab initio computations from $^{78}$Ni towards $^{70}$Ca along neutron number $N=50$}

\author{B. S. Hu} 
\affiliation{Physics Division, Oak Ridge National Laboratory, Oak Ridge, Tennessee 37831, USA}
\affiliation{National Center for Computational Sciences, Oak Ridge National Laboratory, Oak Ridge, Tennessee 37831, USA}

\author{Z. H. Sun} 
\affiliation{Physics Division, Oak Ridge National Laboratory, Oak Ridge, Tennessee 37831, USA}
\affiliation{Department of Physics and Astronomy, Louisiana State University, Baton Rouge, Louisiana 70803, USA}

\author{G. Hagen} 
\affiliation{Physics Division, Oak Ridge National Laboratory, Oak Ridge, Tennessee 37831, USA}
\affiliation{Department of Physics and Astronomy, University of Tennessee, Knoxville, Tennessee 37996, USA}

\author{G.~R. Jansen}
\affiliation{National Center for Computational Sciences, Oak Ridge National Laboratory, Oak Ridge, Tennessee 37831, USA}
\affiliation{Physics Division, Oak Ridge National Laboratory, Oak Ridge, Tennessee 37831, USA}

\author{T. Papenbrock} 
\affiliation{Department of Physics and Astronomy, University of Tennessee, Knoxville, Tennessee 37996, USA}
\affiliation{Physics Division, Oak Ridge National Laboratory, Oak Ridge, Tennessee 37831, USA}


\begin{abstract}
We present coupled-cluster computations of nuclei with neutron number $N=50$ ``south'' of $^{78}$Ni using nucleon-nucleon and three-nucleon forces from chiral effective field theory. We find an erosion of the magic number $N=50$ toward $^{70}$Ca manifesting itself by an onset of deformation and increased complexity in the ground states. For \A{78}Ni, we predict a low-lying rotational band consistent with recent data, which up until now has been a challenge for ab initio nuclear models. 
Ground states are deformed in \A{76} Fe, \A{74} Cr, and \A{72} Ti, although the spherical states are too close in energy to unambiguously identify the shape of the ground state within the uncertainty estimates. In \A{70}Ca, the potential energy landscape from quadrupole-constrained Hartree-Fock computations flattens, and the deformation becomes less rigid. We also compute the low-lying spectra and \BE2 values for these neutron-rich $N=50$ nuclei. 
\end{abstract}
\maketitle

\section{Introduction} 
The neutron-rich nucleus \A{78}{Ni} ($Z=28$, $N=50$) is particularly interesting because it is doubly magic, exhibits shape coexistence~\cite{heyde2011}, and sits at the border of a ``fifth'' island of inversion~\cite{nowacki2016}. Three years before \citet{taniuchi2019} established this experimentally, shape coexistence was already inferred from experiments in neighboring nuclei~\cite{gottardo2016,yangxf2016} and from shell-model calculations~\cite{nowacki2016}, while ab initio computations~\cite{hagen2016b} accurately predicted relatively high energy for the first (spherical) $2^+$ state, consistent with a doubly magic structure. Additional evidence for shape coexistence in the \A{78}{Ni} region has recently been found through high-precision mass measurement \cite{nies2023} in $^{79}_{30}$Zn$^{}_{49}$.

Shell model calculations of \A{78}{Ni} have predicted the low-lying $0^+_2$ state associated with the deformed band to be located at about 2.6~MeV \cite{nowacki2016}. \citet{taniuchi2019} measured the first excited $2_1^+$ state of \A{78}{Ni} at 2.6~MeV, while a state at 2.9~MeV was tentatively assigned to a second deformed $2_2^+$ state, indicating a competition of spherical and deformed configurations. Although the energy of the deformed band's head ($0^+_2$) has not been measured yet, it is expected to be lower than 2.9~MeV. This can be compared to $0^+_2$ excitation energies in other doubly magic nuclei: 6.05~MeV for \A{16}{O} ($Z=N=8$), 3.35~MeV for \A{40}{Ca} ($Z=N=20$), 4.28~MeV for \A{48}{Ca} ($Z=20$, $N=28$), and 3.96~MeV for \A{56}{Ni} ($Z=N=28$). Notably, for \A{68}{Ni} ($Z=28$, $N=40$), which also exhibits shape coexistence \cite{langanke2003,Tsunoda2014,Olaizola2017,nowacki2021}, this energy is only 1.77~MeV. 

As one moves from $^{78}$Ni to lighter neutron-rich elements, spectroscopic data becomes scarce~\cite{nowacki2021}. In iron ($Z=26$) and chromium ($Z=24$), spectroscopic information is known up to $^{72}$Fe and $^{66}$Cr, respectively~\cite{santamaria2015}, while for titanium ($Z=22$) it is only known up to $^{60}$Ti~\cite{gade2014}. However, existing data and shell-model computations indicate that iron and chromium nuclei are deformed between neutron numbers $N=40$ and 50~\cite{nowacki2021}. 
These studies identify \A{74}Cr as the first $N=50$ isotope within the ``fifth'' island of inversion below \A{78}{Ni}, featuring a well-developed deformed ground-state rotational band. While the island of inversion has a simple interpretation in the Nilsson model~\cite{caurier2014,macchiavelli2017}, it is more complicated in the spherical shell model. It occurs when configurations formed by exciting neutrons across the Fermi-level gap~\cite{poves1987,warburton1990,nowacki2021} are more bound than configurations corresponding to neutron magic shell closures. This phenomenon results in a group of nuclei, naively expected to be (near-)spherical in their ground state, becoming deformed. As more neutron-rich isotopes have been studied, islands of inversion around the neutron magic shell at $N=8,20,28,40$ have been recognized on the nuclear chart~\cite{nowacki2021}. 

There are also shell-model calculations~\cite{nowacki2016} of \A{70}Ca, but reliable predictions for this nucleus are difficult to make as continuum effects are expected to be important~\cite{hagen2012b,hagen2016}. Relativistic mean-field computations~\cite{meng2002}, for instance, predict a bunching of single-particle orbitals at the Fermi surface as more and more neutrons are added in calcium beyond $N=40$. These results, together with mean-field computations \cite{neufcort2019} and the discovery of \A{60}Ca \cite{tarasov2018}, suggest that the drip line in calcium could stretch out much beyond $N=40$ and that the structure of \A{70}Ca might be finely tuned and complicated.  

This paper presents {\it ab initio} computations of $N=50$ nuclei south of \A{78}Ni. Such calculations have advanced tremendously in recent years, moving from nuclei near shell closures~\cite{morris2018,gysbers2019,arthuis2020,hu2022} to open-shell systems~\cite{yao2020,stroberg2021,hagen2022,Frosini:2021sxj,sun2024,hu2024}, see, e.g., Ref.~\cite{hergert2020} for a recent review. We employ the coupled-cluster method following Refs.~\cite{sun2024,hu2024}. Those computations captured short-range correlations via particle-hole excitations of a deformed reference state and collective correlations by angular-momentum projection. We will compare our results with the shell-model computations of \citet{nowacki2016} and \citet{taniuchi2019} and the recent valence-space density matrix renormalization group (VS-DMRG) computations of ~\citet{tichai2024}. In our computations, we focus on spectra and electromagnetic transitions.

\section{Methods}
We start from the intrinsic Hamiltonian
\begin{equation}
H=\sum_{i=1}^A\left(1-\frac{1}{A}\right) \frac{\boldsymbol{p}_i^2}{2 m}+\sum_{i<j}^A\left(V_{i j}^{\mathrm{NN}}-\frac{\boldsymbol{p}_i \cdot \boldsymbol{p}_j}{m A}\right)+\sum_{i<j<k}^A V_{i j k}^{3\mathrm{N}},
\end{equation}
where $\boldsymbol{p}$ is the nucleon momentum in the laboratory frame, $m$ the nucleon mass, $V^{\rm NN}$ the nucleon-nucleon interaction, and $V^{\rm 3N}$ the three-nucleon interaction. We use the chiral 1.8/2.0(EM) interaction \cite{hebeler2011}, which yields accurate ground-state energies and spectra of light, medium, and even heavy mass nuclei~\cite{hagen2016b,simonis2017,morris2018,gysbers2019,stroberg2021,hu2022b,hebeler2023}. In practical calculations, the nucleon-nucleon interaction is expressed in the harmonic-oscillator basis with spacing $\hbar\omega$ and single-particle energies up to $(N_{\rm max}+3/2)\hbar\omega$; the three-nucleon interaction is truncated to excitation energies of three nucleons up to $E_{\rm 3max}=28\hbar \omega$~\cite{miyagi2023}. This is sufficient for converged energies, see Fig.~\ref{Ni78_e3max} of the Supplementary Material.
To overcome the computational challenges posed by the large number of three-nucleon matrix elements, we use the normal-ordered two-body approximation~\cite{hagen2007a,roth2012,djarv2021}, modified for deformed nuclei so that rotational invariance of the Hamiltonian is preserved~\cite{Frosini:2021tuj}, while 
 effectively capturing the contributions from three-nucleon forces.

Using this Hamiltonian, we employ axially-symmetric Hartree-Fock calculations with the Hamiltonian $H'=H-\lambda Q_{20}$, where $Q_{20}$ is the mass quadrupole operator and $\lambda$ is a Lagrange multiplier. Following the augmented Lagrangian method of Ref.~\cite{staszscak2010}, we adjust $\lambda$ to achieve the desired quadrupole deformation $\langle Q_{20}\rangle$. This procedure allows us to map out a potential energy surface, defined as the Hartree-Fock energy of the Hamiltonian $H$ at a given deformation $\langle Q_{20}\rangle$. Each local minimum on this energy surface represents a distinct deformed configuration, with the corresponding Slater determinant serving as the reference state for subsequent coupled-cluster computations.

We then perform coupled-cluster with singles and doubles (CCSD) computations~\cite{kuemmel1978,bartlett2007,shavittbartlett2009,hagen2014} and apply symmetry projection~\cite{hagen2022,sun2024} to incorporate both short-range (dynamical) and long-range (static) correlations~\cite{sun2024}. The symmetry projection naturally yields rotational bands. We also compute electromagnetic transition strengths \BE2~$\equiv B({\rm E2},2^+\to 0^+)$, as described in Ref.~\cite{sun2024} and applied to nuclei in the $A\sim 80$ mass region ~\cite{hu2024}.
In what follows, we compare our results with data (available only in \A{78}Ni) and to results from shell-model computations~\cite{nowacki2016,taniuchi2019} and the VS-DMRG~\cite{tichai2024}.

\section{Results and Discussion}
\begin{figure}
\setlength{\abovecaptionskip}{0pt}
\setlength{\belowcaptionskip}{0pt}
\includegraphics[scale=0.68]{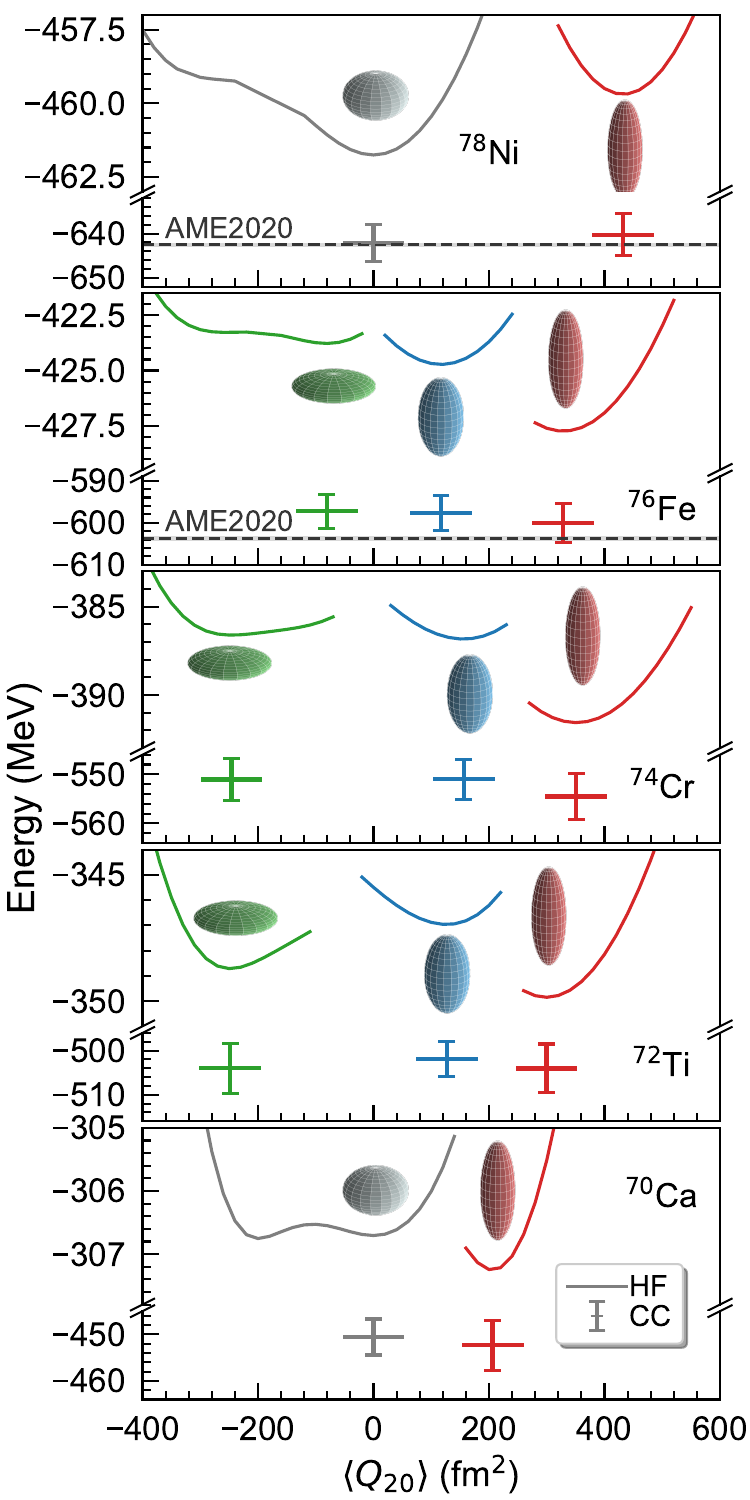}
\caption{\label{energy_N50} Quadrupole constrained energies for states with different shapes (green, gray, blue, and red lines and symbols show oblate, spherical, prolate, and larger prolate configurations, respectively) of the $N=50$ isotones \A{78}Ni, \A{76}Fe, \A{74}Cr, \A{72}Ti, and \A{70}Ca (from top to bottom). The solid line shows the unprojected Hartree-Fock energy surface calculated within model space $N_{\rm max}$=10. Coupled-cluster (CC) results within $N_{\rm max}$=12 that include estimated energy contributions from triples excitations and angular-momentum projection are shown in the lower part for the corresponding Hartree-Fock minima; the uncertainty estimates reflect model-space truncations. The horizontal gray dashed line shows the atomic mass evaluation of 2020 (AME2020)~\cite{Wang2021}.}
\end{figure}

Figure~\ref{energy_N50} shows ground-state computations of the $N=50$ nuclei \A{78}Ni, \A{76}Fe, \A{74}Cr, \A{72}Ti, and \A{70}Ca (from top to bottom). For each nucleus, the upper part of its panel shows the unprojected Hartree-Fock potential energy surfaces in the vicinities of the oblate, spherical, prolate, or larger prolate minima, each shape color-coded appropriately. The Hartree-Fock computations employed eleven major oscillator shells ($N_{\rm max}=10$) with the oscillator frequency $\hbar \omega = 12$~MeV. This model space is sufficiently large to achieve converged results within a few hundred keV~\cite{hu2024}. 
Shape coexistence is clearly visible.
The spherical states with $\langle Q_{20}\rangle=0$ in \A{78}{Ni} and \A{70}{Ca} result from the harmonic-oscillator shell closures at \{$Z=20$, $N=50$\} and \{$Z=28$, $N=50$\}, respectively. 
The oblate configurations with $\langle Q_{20}\rangle<0$  for all calculated nuclei are formed by filling the occupied proton $1f_{7/2}$ shell at the Fermi surface from high to low values of $|j_{z}|$, while maintaining the $N=50$ shell closure for the neutrons. The prolate configurations with $\langle Q_{20}\rangle>0$ in \A{76}{Fe}, \A{74}{Cr}, and \A{72}{Ti} are similarly obtained by filling the low values of $|j_{z}|$ first. The larger prolate configuration in \A{76}{Fe}, \A{74}{Cr}, and \A{72}{Ti}, as well as the prolate configuration in $^{70}$Ca, primarily result from the inversion of \{1$g_{9/2, \ \pm 9/2}$ $\leftrightarrow$ 2$d_{5/2, \ \pm 1/2}$\} for neutrons, with protons filling $1f_{7/2}$ shell from low to high values of $|j_{z}|$ at the Fermi surface. The prolate configuration in $^{78}$Ni is obtained from the inversion of proton \{1$f_{7/2, \ \pm 7/2}$ $\leftrightarrow$ 1$f_{5/2, \ \pm 1/2}$\} and neutron\{1$g_{9/2, \ \pm 9/2}$ $\leftrightarrow$ 2$d_{5/2, \ \pm 1/2}$\} at the Fermi surface. 

The prolate minima in \A{78}{Ni} and \A{70}{Ca} have deformation parameters of $\beta_2 \approx 0.28$ and 0.16, respectively. The larger prolate minima in \A{76}{Fe}, \A{74}{Cr}, and \A{72}{Ti} have $\beta_2$ values of approximately 0.22, 0.25, and 0.22, respectively. Here, $\beta_2$ is calculated from the mass quadrupole moment $\langle Q_{20}\rangle$ using the formula $\beta_2 \equiv \sqrt{5 \pi}\left\langle Q_{20}\right\rangle/(3 A R_0^2)$, with $R_0=1.2A^{1/3}$~fm~\cite{pritychenko2016}. It is interesting that the shapes of $^{78}$Ni and $^{74}$Cr at the local Hartree-Fock minima, computed with a chiral interaction, are consistent with the calculations using the shell-model PFSDG-U interaction~\cite{nowacki2016}, although we obtained smaller $\beta_2$ values for $^{76}$Fe.

The lower part of each nucleus's panel in Fig.~\ref{energy_N50} shows coupled cluster results. Here, the deformed Hartree-Fock state served as a reference at each minimum. We computed energies in the CCSD approximation and added estimated energy contributions from triples excitations and angular momentum projection as follows: The effect of triples excitations on the ground-state energies is estimated as 10\% of the CCSD correlation energy~\cite{bartlett2007,sun2022,Ekstrom:2022yea}. The unprojected coupled-cluster computations employed a model-space of $N_{\rm max} = 12$ with the oscillator frequency varied in the range $\hbar \omega = 10, 12, 14, 16$~MeV. The final coupled-cluster result shown in Fig.~\ref{energy_N50} used the frequencies $\hbar \omega = 12$ and 14~MeV for different shapes in the computed nuclei. The (small) contribution from angular-momentum projection is obtained in CCSD from a smaller model space with $N_{\rm max} = 6$. This is sufficient to capture the corresponding static correlations~\cite{hu2024}. The uncertainties are from comparing differences between unprojected CCSD results at $N_{\rm max} = 10$ and $N_{\rm max} = 12$. The coupled cluster results for \A{78}{Ni} and \A{76}{Fe} are consistent with the extrapolated masses from the atomic mass evaluation of 2020 (AME2020)~\cite{Wang2021}. Details about the convergence of our results with respect to the employed model spaces are shown in Fig.~\ref{Energy_CC_HF} of the Supplementary Material.

The coupled-cluster computations for $^{78}$Ni show that the spherical configuration has the lowest energy for the employed model spaces ($N_{\rm max}=10-12$ and $\hbar\omega=12-16$ MeV). Similarly, the deformed shapes have the lowest energy in \A{76}Fe and \A{74}Cr. However, the uncertainties shown in Fig.~\ref{energy_N50} reflect uncertainties of the ground-state shape. For \A{72}{Ti} and \A{70}{Ca}, different model spaces (varying $N_{\rm max}$ and $\hbar\omega$) yield both positive and negative energy gaps between spherical and prolate shapes, making it challenging to unambiguously identify the ground-state shape. The nucleus $^{70}$Ca exhibits no well-defined minima. This entails considerable uncertainties in its structure, consistent with Refs.~\cite{meng2002,neufcort2019,stroberg2021}. For this reason, we refrain from computing spectra and the \BE2 transition strength in this nucleus. A credible computation would require us to use generator coordinate methods and to include coupling to the continuum. This is beyond our present capabilities. 

While the results shown in Fig.~\ref{energy_N50} demonstrate shape coexistence in the $N=50$ nuclei, they come with caveats due to the following limitations of our computations. 
(i) We neglected continuum effects which are certainly important in \A{70}{Ca} and possibly also for the excitation spectrum of \A{72}Ti. (ii) We are limited to axial symmetry and can not capture any static triaxial deformation or $\gamma$-softness. (iii) The angular-momentum projection is done after variation. This is accurate for rigidly deformed states but less so in cases where one needs to capture the mixing and superposition of configurations with different deformations.  

\begin{figure*}
\setlength{\abovecaptionskip}{0pt}
\setlength{\belowcaptionskip}{0pt}
\includegraphics[scale=0.50]{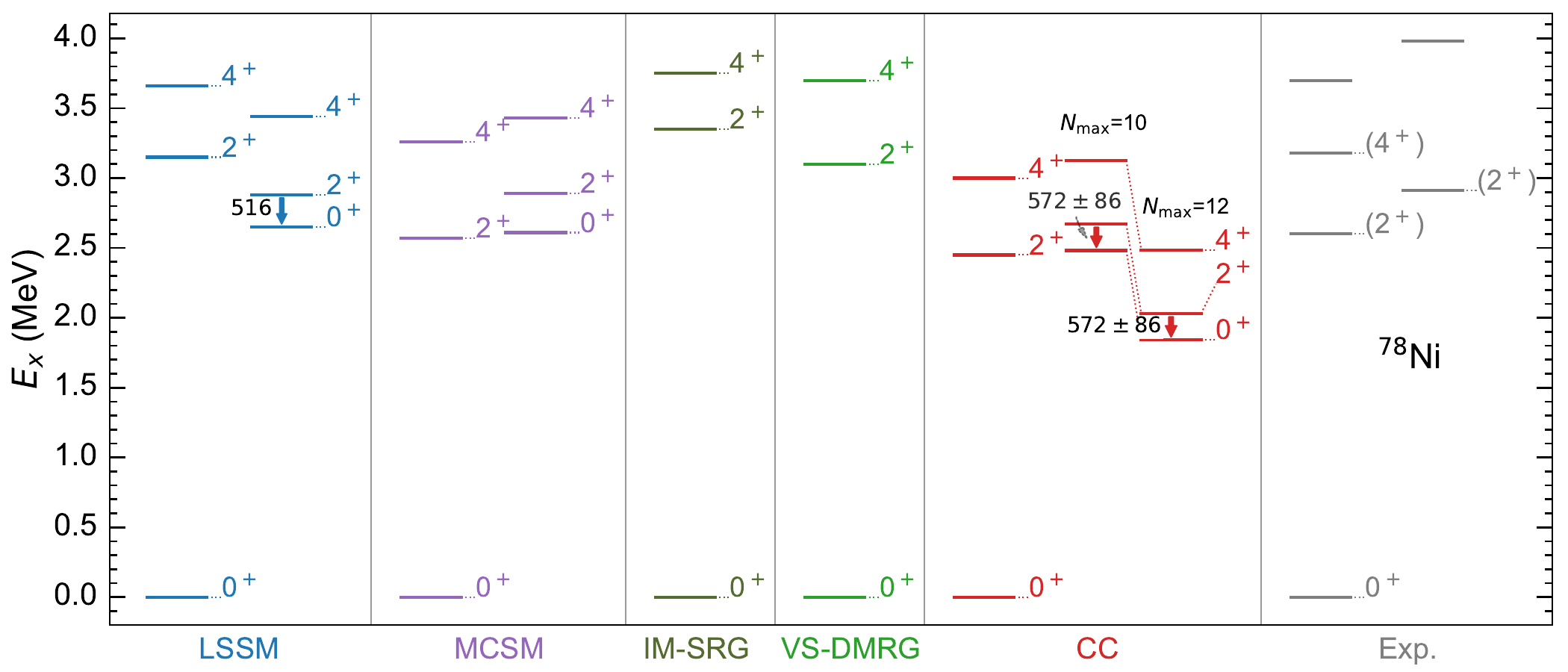}
\caption{\label{spectra_Ni78} Comparison of theoretical calculations with experimental data~\cite{taniuchi2019} for low-lying levels in $^{78}$Ni. Theoretical results include the large-scale shell model (LSSM)~\cite{nowacki2016} and Monte Carlo shell model (MCSM)~\cite{taniuchi2019} using more phenomenological interactions, the valence-space formulation~\cite{stroberg2017} of the in-medium similarity renormalization group (IM-SRG)~\cite{taniuchi2019}, the same IM-SRG solved with the VS-DMRG~\cite{tichai2024}, and coupled-cluster (CC) results for the spherical shape~\cite{hagen2016b} and for the prolate shape from this work. Model-space uncertainties in prolate states are shown using different $N_{\rm max}$ values labeled nearby. $B({\rm E}2;2^+ \rightarrow 0^+_2)$ values (in units of $e^2$fm$^4$) are also shown.}
\end{figure*}

Figure~\ref{spectra_Ni78} displays the spectra obtained from coupled-cluster calculations of \A{78}{Ni} and compares them with experimental data \cite{taniuchi2019} and theoretical results from shell-model calculations, in-medium similarity renormalization group (IM-SRG), and VS-DMRG. The data is close to the large-scale shell-model  predictions by \citet{nowacki2016} and the Monte Carlo shell-model results of \citet{taniuchi2019}. Unfortunately, no data is available for the $0^+_2$ state, which is predicted to be the head of the deformed band. The coupled-cluster computations of the ground-state band were obtained from the equation-of-motion formalism excited from the spherical ground state and taken from Ref.~\cite{hagen2016b}. 
The coupled-cluster computations of the deformed band employed a deformed reference state in model spaces with $N_{\rm max}=10$ and 12, and $\hbar\omega=14$~MeV; the difference between the two are used to establish model-space uncertainties. The angular-momentum projection in coupled-cluster theory, which yields rotational bands, is computationally expensive and restricted to $N_{\rm max}=6$ for spectra and $N_{\rm max}=8$ for the \BE2's in this work. Therefore, we added the energy gain from projection in $N_{\rm max}=6$ to the unprojected coupled-cluster results in $N_{\rm max}=10$ and 12.
This is justified because collective rotational phenomena are mainly related to long-range physics and $N_{\rm max}=6$, 8  model spaces are large enough for the computation of rotational bands~\cite{sun2024} and \BE2 calculations~\cite{hu2024}. Despite the significant uncertainty in the excitation energy of band head $0^+_2$, the rotational bands exhibit a similar pattern as those of the shell-model calculations.

The results from a valence-space formulation of the IM-SRG~\cite{taniuchi2019} and from the VS-DMRG~\cite{tichai2024} used the same interaction as the present coupled-cluster calculations. In those approaches, a valence-space (i.e. shell-model) interaction is generated via the IM-SRG method~\cite{stroberg2017}. The depicted IM-SRG results are from an approximate diagonalization in the valence space, while the VS-DMRG uses the density matrix renormalization group~\cite{white1992} for a more accurate solution. 
The common valence-space interaction from the IM-SRG was generated with a truncation at the normal-ordered two-body level, presumably leading to the inflated spherical spectrum~\cite{taniuchi2019}. Truncations at the normal-ordered three-body level have been developed~\cite{heinz2021,stroberg2024} and they yield accurate energies for $2^+$ states in carbon nuclei~\cite{stroberg2024b}.

\begin{figure*}
\setlength{\abovecaptionskip}{0pt}
\setlength{\belowcaptionskip}{0pt}
\includegraphics[scale=0.50]{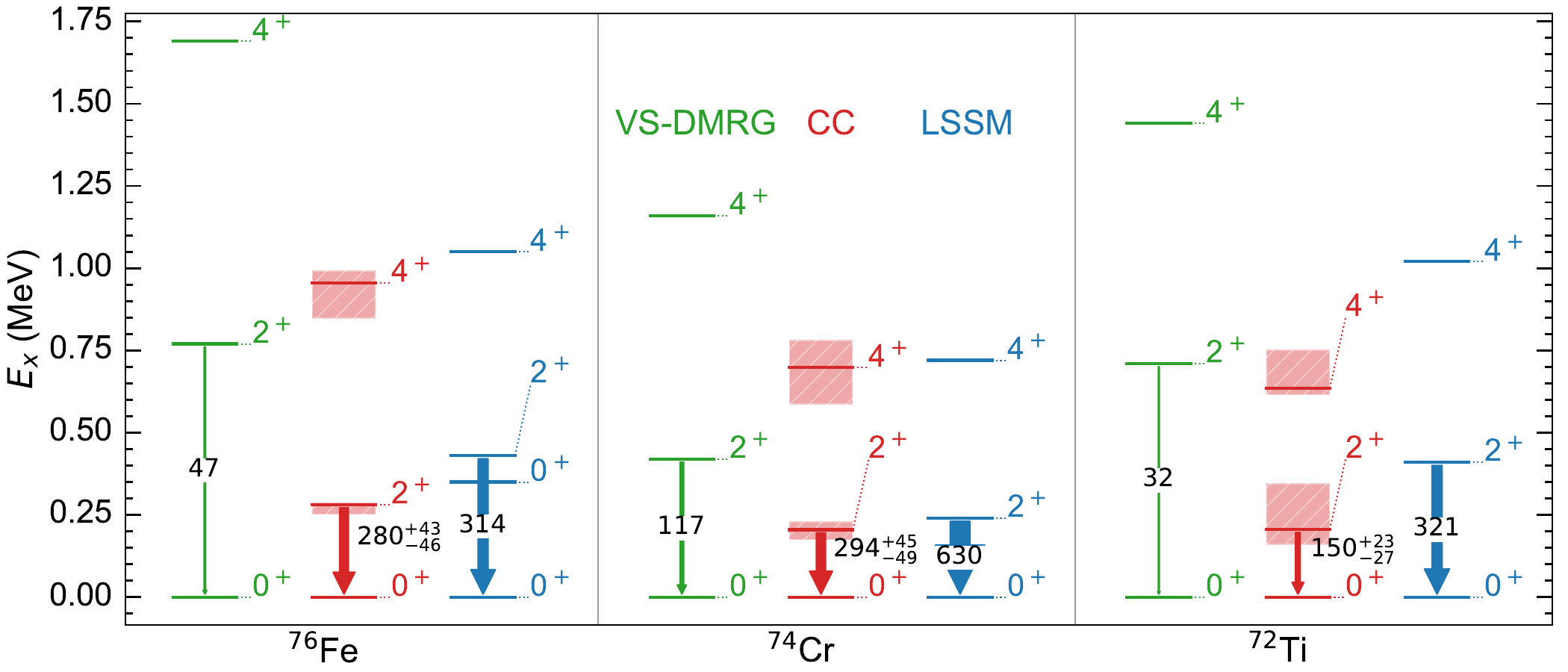}
\caption{\label{spectra_N50} Excitation energies of $^{76}$Fe, $^{74}$Cr, $^{72}$Ti and $^{70}$Ca from the valence-space density matrix renormalization group (VS-DMRG)~\cite{tichai2024}, projected coupled cluster (CC) of this work, and large-scale shell model (LSSM)~\cite{nowacki2016}. The uncertainty estimates of the coupled-cluster results are based on varying the harmonic oscillator frequency $\hbar\omega$ from 10 to 16~MeV and a 15\% method error as done in Ref.~\cite{sun2024}. $B({\rm E}2;2^+_1 \rightarrow 0^+_1)$ values (in units of $e^2$fm$^4$) are also indicated.}
\end{figure*}

Predictions of excitation energies in $^{76}$Fe, $^{74}$Cr, and $^{72}$Ti are presented in Fig.~\ref{spectra_N50}.  \citet{nowacki2016} identified three minima -- spherical, oblate, and prolate --  in \A{78}{Ni} and \A{76}{Fe} using constrained Hartree-Fock calculations, while \A{74}{Cr} was dominated by a single prolate minimum. They proposed that the islands of inversion at $N=40$ and $N=50$ merge around the chromium isotopes and that the $N=50$ island of inversion is located below $^{78}$Ni. These findings are also supported by the shell-model analysis of Ref.~\cite{li2023}. Coupled-cluster computations reveal prolate states with larger deformation in $^{76}$Fe, $^{74}$Cr, and $^{72}$Ti, arising from an intruder configuration built by promoting neutrons across the $N=50$ shell. These specific levels are shown in Fig.~\ref{spectra_N50} for the coupled-cluster  results.
They are similar to the shell-model results  of~\citet{nowacki2016}. In contrast, The VS-DMRG spectra from ~\citet{tichai2024} are inflated compared to those from the shell model and coupled-cluster theory.   

\begin{figure}
\setlength{\abovecaptionskip}{0pt}
\setlength{\belowcaptionskip}{0pt}
\includegraphics[scale=0.38]{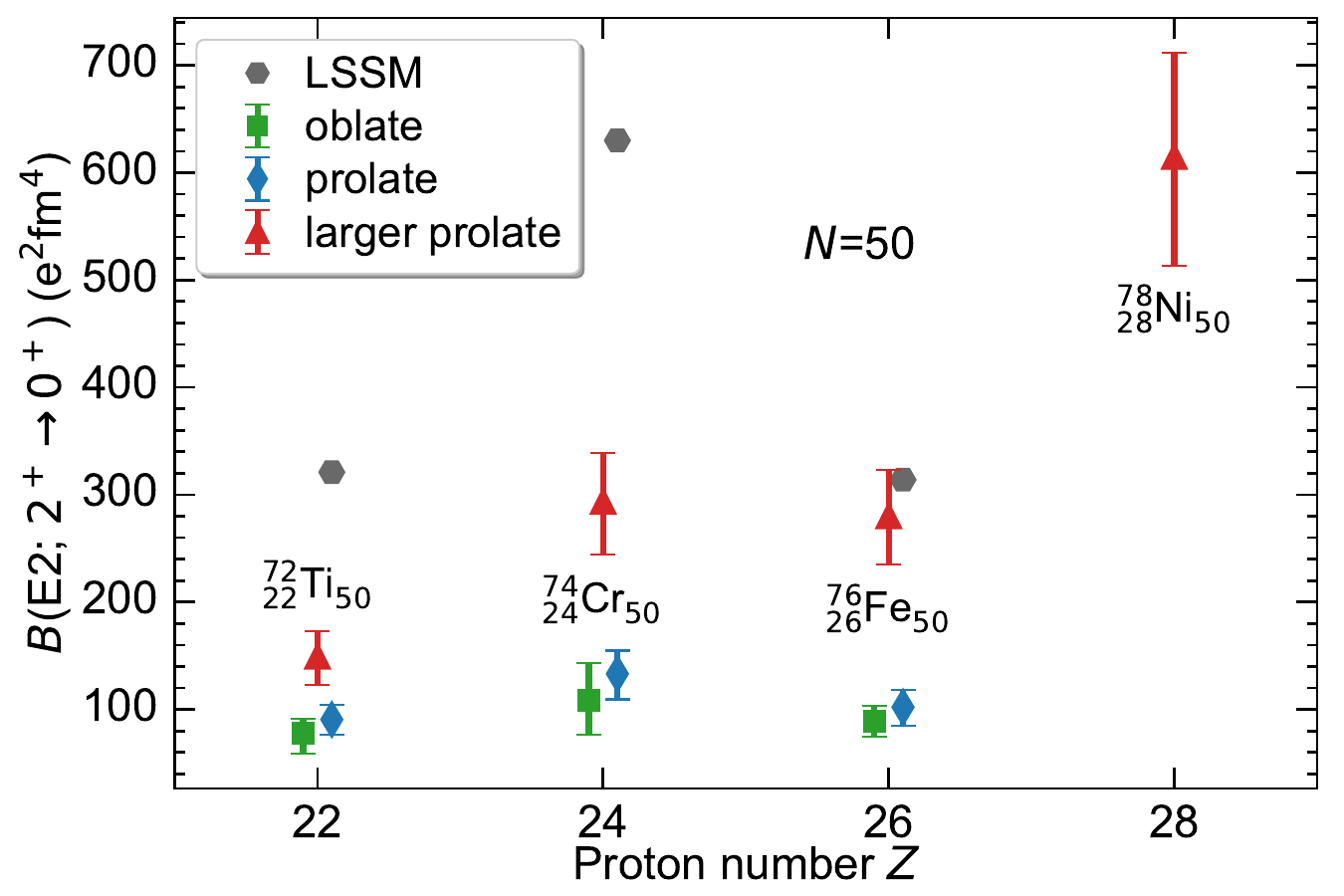}
\caption{\label{BE2_N50} $B({\rm E}2;2^+_1 \rightarrow 0^+_1)$ values of $^{72}$Ti, $^{74}$Cr, $^{76}$Fe and $^{78}$Ni from the large-scale shell model (LSSM)~\cite{nowacki2016} and the projected coupled-cluster (CC) method of this work. The uncertainty of CC is estimated based on varying the harmonic oscillator frequency $\hbar\omega$ from 10 to 14~MeV and 15\% method error.}
\end{figure}

Let us turn to electromagnetic transition strengths \BE2. 
We note that the coupled-cluster \BE2 values are from intra-band transitions, i.e., between states obtained from angular-momentum projection of the same deformed configuration. We currently lack the capability to compute inter-band transitions. In contrast to the shell model the coupled-cluster and VS-DMRG methods do not employ effective charges. 
Uncertainty estimates for \BE2 transition strengths are based on varying the harmonic oscillator frequency $\hbar\omega$ from 10 to 14~MeV and on a 15\% method error that was estimated in Ref.~\cite{sun2024} from benchmarks with the symmetry-adapted no-core shell model~\cite{dytrych2020,launey2020}. Results for \A{78}Ni are shown in Fig.~\ref{spectra_Ni78}. Here the coupled-cluster results for the transition in the deformed band agree with shell-model results within uncertainties. Figure~\ref{BE2_N50} shows the corresponding results for \A{76}Fe, \A{74}Cr, and \A{72}Ti. The coupled-cluster results align with the shell-model results  in \A{76}Fe, but the latter is about a factor two larger for \A{74}Cr and \A{72}Ti. The VS-DMRG results are typically much smaller than the coupled-cluster results.
Measurements are needed to clarify this situation. 

\section{Summary}
We investigated shell evolution, shape coexistence, and collectivity as one moves from  $^{78}$Ni towards $^{70}$Ca along the neutron number $N=50$ line. Our computations use the coupled-cluster method and accurate interactions from chiral effective field theory. We find that the magic neutron number $N=50$ erodes south of $^{78}$Ni with an onset of deformation appearing in the ground states. We reproduce known levels in \A{78}Ni from first principles and, in particular, predict a low-lying rotational band consistent with data, which has been a challenge for ab initio nuclear modeling. We also studied how deformed configurations evolve towards \A{70}Ca and confirm that accurate computations of this nucleus are challenging. We find that spectra of \A{76}Fe, \A{74}Cr, and \A{72}Ti largely agree with shell-model results within our uncertainty estimates, though \A{72}Ti is found to be more rotational. In contrast, our \BE2 values are smaller for nuclei lighter than iron when compared to results obtained with the shell model.

\section*{Acknowledgements}
We thank Takayuki Miyagi for the {\tt NuHamil} code~\cite{miyagi2023} and Ragnar Stroberg for the {\tt imsrg++} code~\cite{Stro17imsrg++} used to generate matrix elements of the chiral three-body interaction. This work was supported by the U.S. Department of Energy (DOE), Office of
Science, under SciDAC-5 (NUCLEI collaboration) and contract DE-FG02-97ER41014, by the Quantum Science Center, a National Quantum Information Science Research Center of the U.S. Department of Energy. Computer time was provided by the Innovative and Novel Computational Impact on Theory and Experiment (INCITE) program. This research used resources from the Oak Ridge Leadership Computing Facility located at ORNL, which is supported by the Office of Science of the Department of Energy under Contract No. DE-AC05-00OR22725.

%


\section*{Supplementary Material}
\begin{figure}
\setlength{\abovecaptionskip}{0pt}
\setlength{\belowcaptionskip}{0pt}
\includegraphics[scale=0.56]{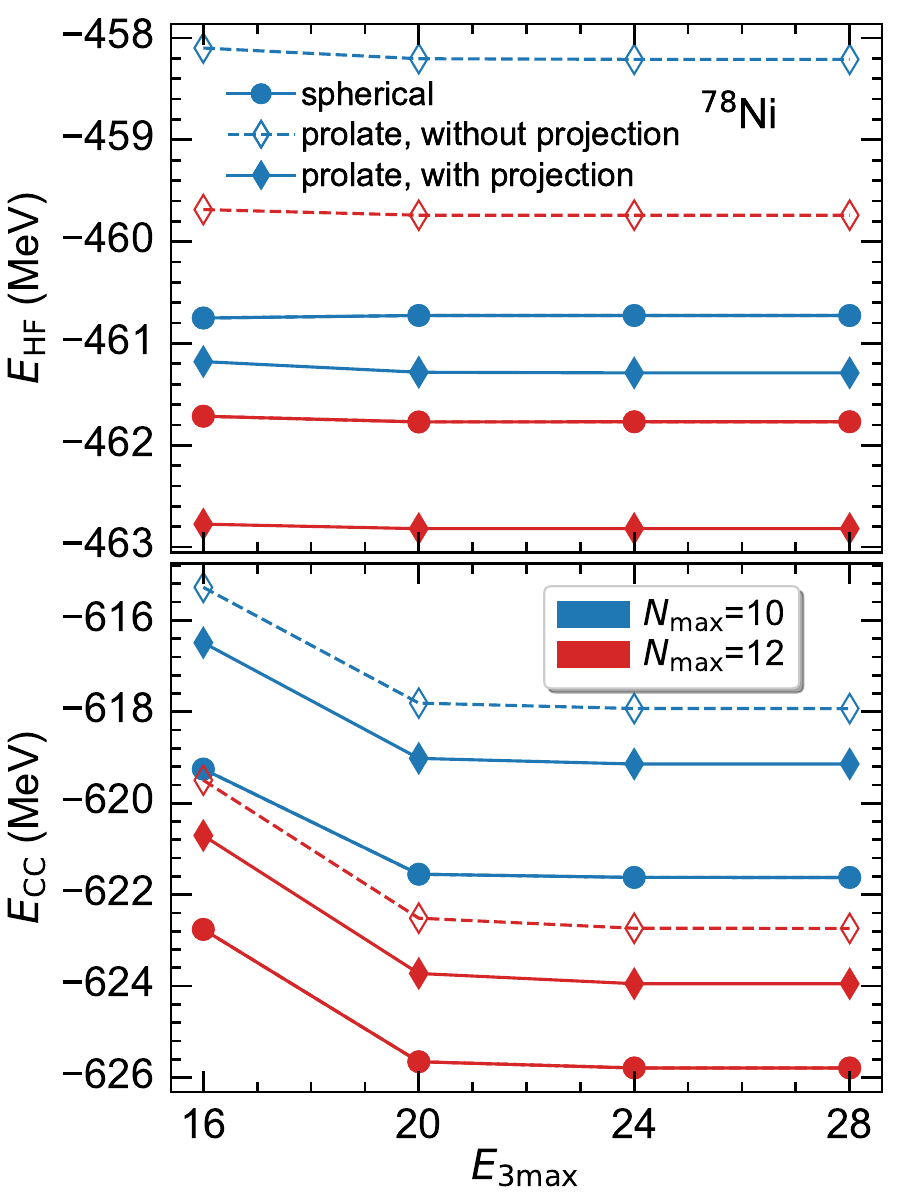}
\caption{\label{Ni78_e3max} Convergence of Hartree-Fock and coupled cluster calculations as a function of three-body truncation $E_{\rm 3max}$.}
\end{figure}

The Supplementary Material details how the results presented in the main text depend on model-space parameters. 
Figure~\ref{Ni78_e3max} shows the dependence of the Hartree-Fock energy ($E_{\rm HF}$) and the energy from coupled-cluster computations ($E_{\rm CC}$) depend on the energy cutoff $E_{\rm 3max}$ employed in the three-body matrix elements. Here, the coupled-cluster energy is based on the axially deformed reference state and uses the coupled-cluster singles and doubles approximation. Results are also shown for different energy cutoffs $E=(N_{\rm max}+3/2)\hbar\omega$ of the employed single-particle basis from the harmonic oscillator. The employed oscillator spacing is $\hbar\omega=14$~MeV. 

\begin{figure*}
\setlength{\abovecaptionskip}{0pt}
\setlength{\belowcaptionskip}{0pt}
\centering
\begin{subfigure}{}
\includegraphics[scale=0.50]{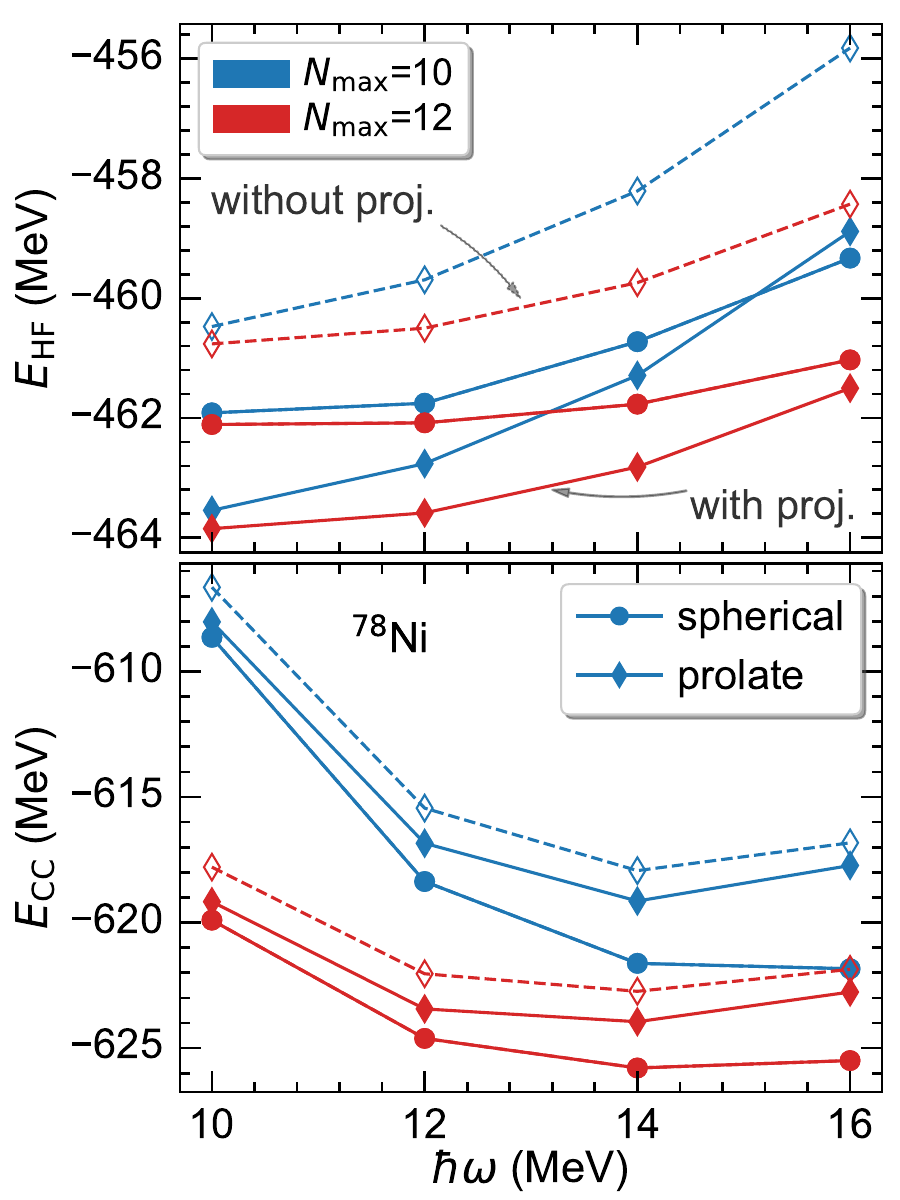}
\end{subfigure}
\begin{subfigure}{}
\includegraphics[scale=0.50]{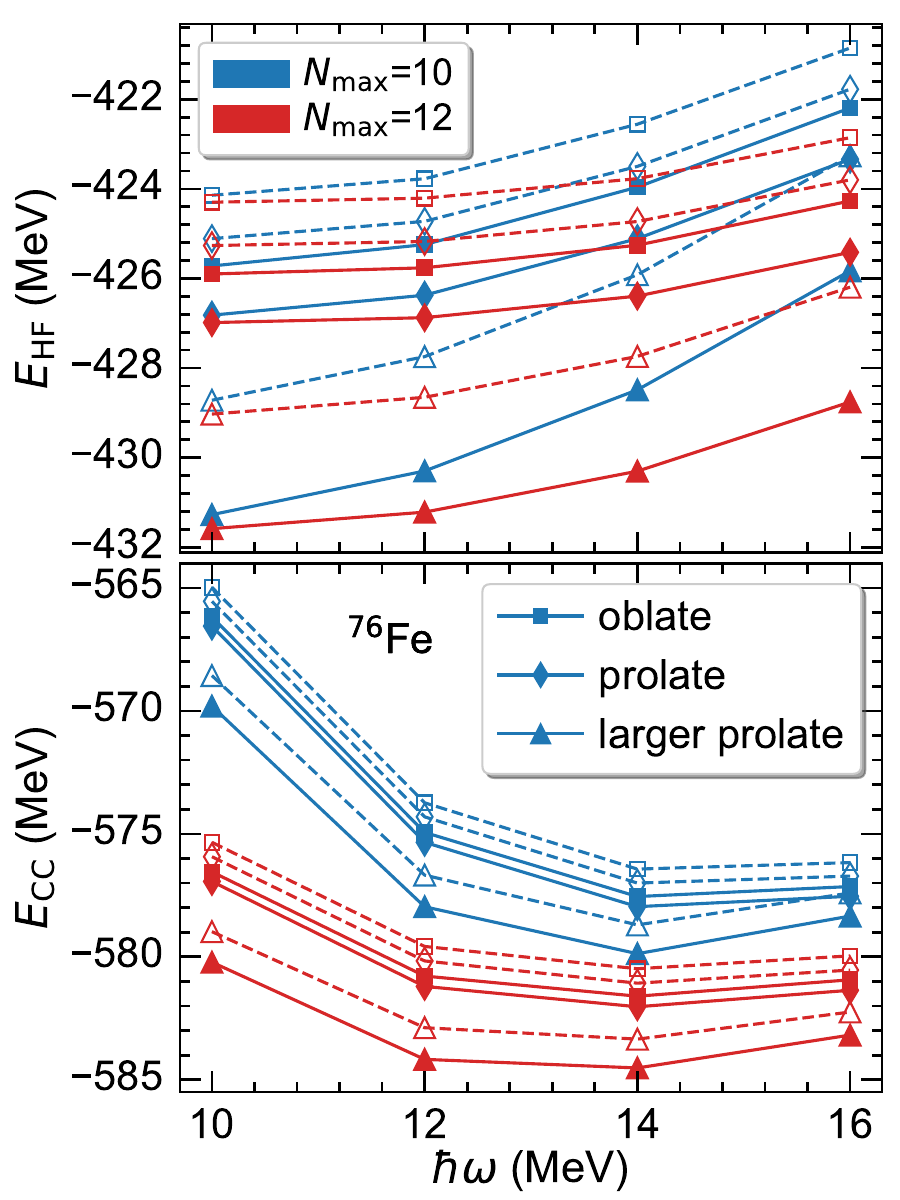}
\end{subfigure}
\\
\begin{subfigure}{}
\includegraphics[scale=0.50]{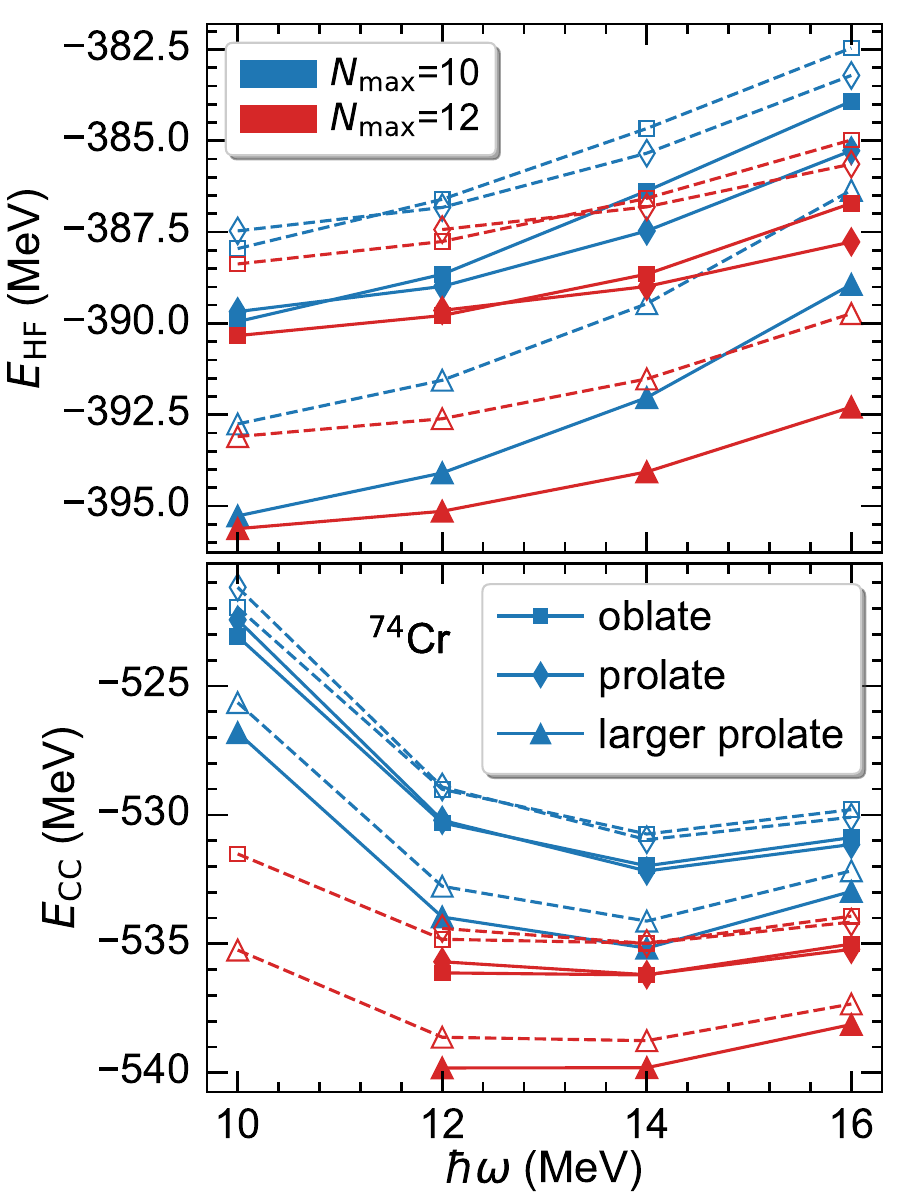}
\end{subfigure}
\begin{subfigure}{}
\includegraphics[scale=0.50]{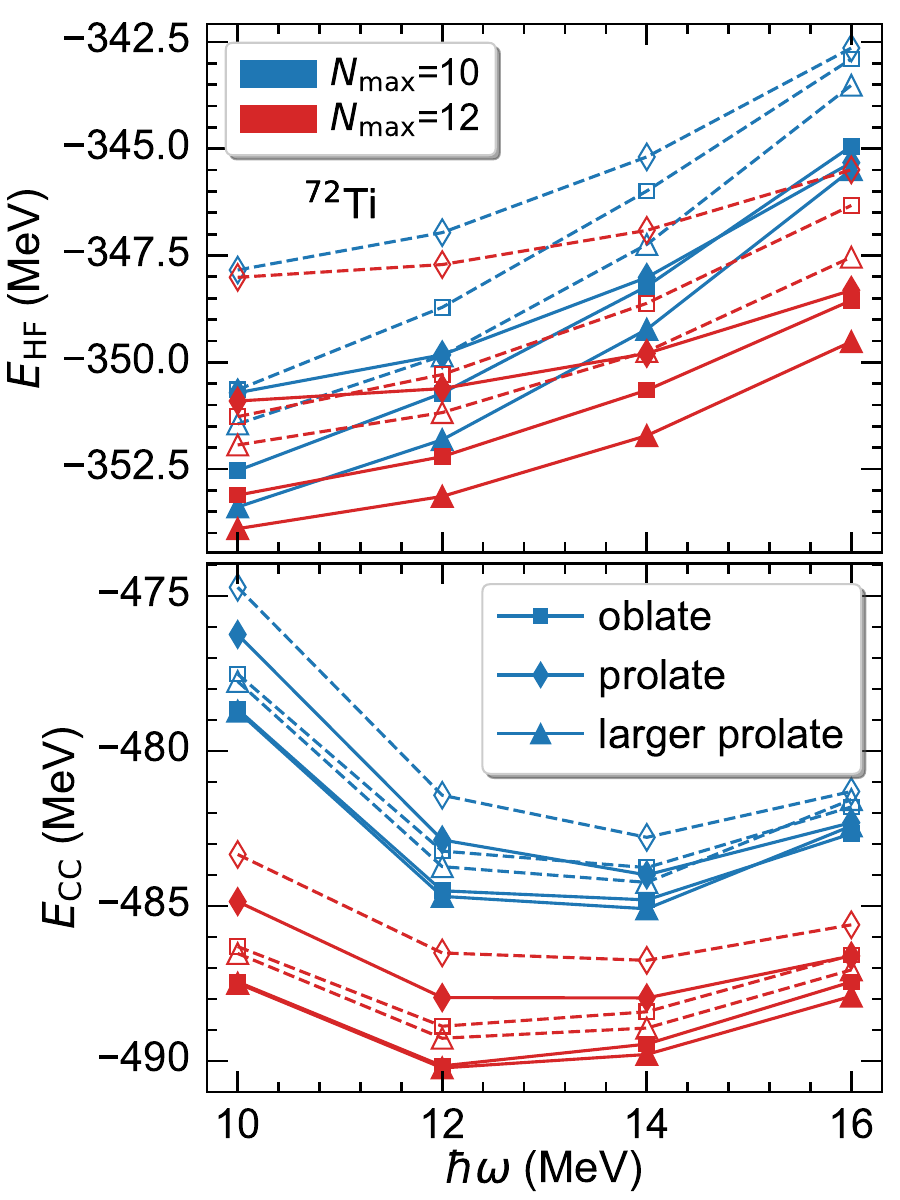}
\end{subfigure}
\caption{\label{Energy_CC_HF} Energies of $^{78}$Ni, $^{76}$Fe, $^{74}$Cr, and $^{72}$Ti (from top to bottom) from Hartree Fock (upper) and coupled cluster (lower) calculations as a function of the harmonic-oscillator frequency $\hbar\omega$ within the model space $N_{\rm max} =10$ and 12. Different colors indicate different deformations. Dashed lines are from symmetry-breaking calculations, while full lines are from symmetry projection. The Hartree-Fock result with projection indicates the projection-after-variation (PAV) Hartree-Fock energy. For the projected coupled cluster results, the contribution from angular momentum projection was obtained from a projected CCSD calculation in a smaller model space with $N_{\rm max} = 6$.}
\end{figure*}

Figure~\ref{Energy_CC_HF} shows the convergence of the Hartree-Fock energies ($E_{\rm HF}$) and the energies from coupled-cluster computations ($E_{\rm CC}$) as a function of the employed harmonic oscillator frequency and for different $N_{\rm max}$ for the nuclei computed in this work. 

\end{document}